\documentclass[lettersize,journal]{IEEEtran}
\usepackage{graphicx} 
\usepackage{threeparttable}
\usepackage{framed}
\usepackage{mdframed}
\usepackage{subcaption}
\usepackage{fontawesome5}
\usepackage{tabularx}
\usepackage{graphicx,calc}
\usepackage{wrapfig}
\usepackage[hidelinks]{hyperref}
\usepackage{booktabs}

\newcommand{\qq}[2]{{\scriptsize \sf ``#1''~{\tiny{(#2)}}}}

\newcommand{\mots}{\includegraphics[scale=0.4, trim={0.2cm 0.1cm 0 0 0},clip]{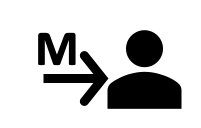}\,}

\newcommand{\ts}
{\includegraphics[scale=0.4, trim={0 0.1cm 0 0 0},clip]{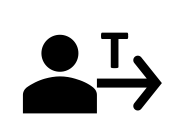}}

\newcommand{\so}
{\includegraphics[scale=0.4, trim={0 0.1cm 0 0 0},clip]{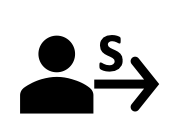}}

\begin{document}

\title{The Software Documentor Mindset}

\author{
\IEEEauthorblockN{Deeksha M. Arya, Jin L.C. Guo, Martin P. Robillard}

\IEEEauthorblockA{McGill University}

\IEEEauthorblockA{\emph{deeksha.arya@mail.mcgill.ca}, \emph{jguo@cs.mcgill.ca}, \emph{robillard@acm.org}}
}

\markboth{The Software Documentor Mindset}{}

\maketitle

\begin{abstract}
Software technologies are used by programmers with diverse backgrounds. To fulfill programmers' need for information, enthusiasts contribute numerous learning resources that vary in style and content, which act as documentation for the corresponding technology. We interviewed 26 volunteer documentation contributors, i.e. \textit{documentors}, to understand why and how they create such documentation. From a qualitative analysis of our interviews, we identified a total of sixteen \textit{considerations} that documentors have during the documentation contribution process, along three \textit{dimensions}, namely \textit{motivations}, \textit{topic selection techniques}, and \textit{styling objectives}. We grouped related considerations based on common underlying themes, to elicit five \textit{software documentor mindsets} that occur during documentation contribution activities. We propose a structure of mindsets, and their associated considerations across the three dimensions, as a framework for reasoning about the documentation contribution process. This framework can inform information seeking as well as documentation creation tools about the context in which documentation was contributed.
\end{abstract}


\section{Introduction}
\label{sec:Introduction}

\IEEEPARstart{W}{ith} the continuous advancement of the software development industry, new tools and technologies are continuously developed, updated, and released. Technology creators publish accompanying software documentation to support programmers who need to understand and use these technologies. Thus, the documentation acts as a knowledge base for technology-related information. Furthermore, enthusiastic programmers and users of the technology have begun to contribute to the documentation landscape in different ways, including via blog articles~\cite{Treude2018} and video tutorials~\cite{Hora2021}. Since documentation is the process of collecting knowledge and ideas~\cite{Briet1951, Lund2007} by converting tacit knowledge to connected, explicit information pieces~\cite{Schneider2009,Robillard2021}, documentation creation is a human-centric process, with the documentation creator at the reins. Consequently, contributing documentation can be tedious~\cite{Shmerlin2015} and time-consuming~\cite{Aghajani2020}. 

One aspect of documentation creation that may be time-consuming is the decision-making process: documentation creators have a variety of decisions to make about content and presentation~\cite{Arya2021}. Different decisions made can result in variations in the \textit{style} of different contributed documentation~\cite{Arya2023}. Consequently, information seekers must use cues~\cite{Piorkowski2015, Arya2022} to search for pertinent information~\cite{Wong2022} that aligns with their preferences~\cite{EarlePreferences, Escobar-Avila2019}. To support this information search process, prior work has focused on improving the efficiency of search~\cite{Jiang2017, ajam2021scout}, or investigating how documentation can be catered to information needs~\cite{Garousi2013, Aghajani2020}. However, understanding the context in which the documentation was created can guide information seekers towards pertinent resources.

We investigated the research question: \textbf{why and how do people voluntarily contribute software documentation online?} We focus on \textit{voluntary documentation creation} and contribute insight on why people share technical information and what the contribution process entails. In our semi-structured interviews with 26 documentation contributors, we asked informants about why they began contributing documentation, and how they go about the creation process. This included, for example, how they determined what technologies and topics to cover and how they made decisions about the style of the documentation. We performed qualitative analysis of the interviews, from which we elicited sixteen \textit{considerations} documentors have, across three major \textit{dimensions} of the documentation contribution process, i.e. \textit{motivations}, \textit{topic selection techniques}, and \textit{styling objectives}. We use the term \textit{documentor} to refer to someone who voluntarily creates and contributes documentation. From our analysis, we  elicited the following considerations: five motivations documentors have for contributing documentation, five techniques documentors used to select topics, and six different objectives they use for styling documentation a particular way.

\IEEEpubidadjcol

We noted that considerations across different dimensions are thematically related. We grouped related considerations and refer to the groups as \textit{software documentor mindsets}. A mindset describes a particular combination of motivations, topic selection techniques, and styling objectives that captures the thought process of the documentor. For example, some informants were motivated to create documentation because of \textit{inadequate documentation}. To select topics, some informants chose ones for which documentation did not exist before, \textit{to fill the documentation gap}. Some informants thought about how \textit{to differ from existing documentation}, when styling their content. From these three considerations, we elicited the mindset: \textit{novelty and value addition}.

The mindsets and their associated considerations provide a framework for characterizing the documentation contribution process. This framework provides context for documentation, which can surface cues for information seekers to use as they search for pertinent resources. For example, knowing that a documentor has the \textit{novelty and value addition} mindset, indicates that their documentation covers information about the technology in an alternate, novel manner. An information seeker struggling to learn a technology can identify that this documentor's documentation will be relevant to them, rather than documentation of another documentor who predominantly has the \textit{growth and visibility} mindset. Similarly, our insights can inform the design of documentation tools to support documentation contribution.
\section{Related Work}
\label{sec:related_work}

We discuss prior research on documentation motivations and creation practices, and mindsets in software engineering.

\subsection{Documentation motivations}
\label{subsec:rel_motivation}

Ryan and Deci proposed the self-determination theory (SDT), which introduced a taxonomy of motivation including amotivation, intrinsic, and extrinsic motivation~\cite{Ryan2020}.
Prior work has also investigated how peoples' motivations affects creativity~\cite{Fischer2019}, knowledge sharing~\cite{Hung2011}, and contribution to open source software~\cite{Gerosa2021}. Personal blogs have been found to be sources of therapeutic reflection and experience sharing~\cite{Gill2009, Xie2016}. Li elicited seven reasons why adults blog, including for \textit{self-expression} and for \textit{socialization}, from a questionnaire filled by 288 bloggers~\cite{Li2005}. Shmerlin et al. conducted interviews with five software developers and a questionnaire with ten developers to understand the motivations of developers to document their code~\cite{Shmerlin2015}. Their participants indicated increased code comprehensibility, structure, and quality as what they enjoyed most about documenting, while acknowledging that it is difficult and time-intensive. McArthur discussed four common prejudices against documentation, one of which is that programmers would rather program than write documentation~\cite{McArthur1986}.

From a survey with thirty bloggers, Parnin et al. reported four types of technical blogging motivations~\cite{Parnin2013}, including for personal branding and as a personal knowledge repository. MacLeod et al. studied screencast documentation wherein developers record their screen and explain how the corresponding technology works~\cite{MacLeod2015}. They analyzed 20 Youtube videos and interviewed ten screencast creators and reported five reasons \textit{why} developers create the screencasts, which included to build an online identity, and to promote themselves. Whereas prior literature has focused on developers' motivations to create either text or video content, our informants include both text and video documentation creators who are not necessarily developers. Additionally, as part of the interview, we encouraged informants to elaborate in detail about how they \textit{began} contributing documentation. We found that the motivations we elicited correspond to the aggregate of those reported by Parnin et al. and Macleod et al., and additionally introduce the novel motivation consideration \textit{related pursuits}.

\subsection{Documentation creation practices}

Bottomley performed interviews with 24 documentation writers to understand their skills and practices~\cite{Bottomly2005}. They found that documentation creation requires skills such as recognizing unclear instructions, understanding high-level concepts, and predicting and responding to user needs. Schmidt proposed an analytical framework for technical blogging that has three structural dimensions: \textit{rules} about the appropriateness and medium of content, \textit{relations} such as hyperlinks to other resources, and the \textit{code} in the blog~\cite{Schmidt2007}. 

Dagenais and Robillard interviewed 22 developers and technical writers who either wrote or read documentation to understand how documentation evolves~\cite{Dagenais2010}. Additionally, they manually inspected the evolution of 19 documents (over 1500 revisions) from 10 open source projects. They reported different decisions that documentation involves, such as determining what kind of documentation, i.e. \textit{getting started} or reference documentation, should be written first. The authors also reported that one way to maintain documentation alongside the evolution of a project is to document the change immediately. Wang et al. reported how the nine identified categories of documentation in 80 popular computational notebooks mapped to stages in the data science life cycle~\cite{Wang2021}. Prior work has elicited how to write articles such that they motivate the reader to read instructions~\cite{Loorbach2007, Karreman2013}. For example, Goodwin provided suggestions for technical writers on how to style manuals such that it ``emplots'', i.e. involves the reader in an action-oriented storyline~\cite{Goodwin1991}. Similarly, different information styles, such as \textit{declarative} and \textit{procedural} information, have varying effects on the instruction-following behaviour of readers~\cite{Ummelen1996,Mackiewicz2005}. Guidelines exist for designing documentation~\cite{Procida_Diataxis} and technical writing~\cite{Bhatti2021}, and techniques exist to support the automatic generation of documentation~\cite{Paris2002,Head2020,Wang2023,Wu2023}. Understanding the thought process of documentors can inform the design of such tools.
 
From a mixed-methods study involving surveys with over 150 developers, and eleven semi-structured interviews, LaToza et al. reported that knowledge about software projects, for example the architecture and design rationale, is not systematically documented~\cite{LaToza2006}. Instead, much of this information is only ``in peoples' heads''. The majority of their participants agreed that understanding the rationale behind code, though important, was challenging, partially because this information was never documented. As a result, there is a need to better support the capture of mental models and design rationale, in order to support decision making~\cite{Heinonen2023} in programming tasks and future stakeholders' understanding of a project.

\subsection{Mindsets in Software Engineering}

There exist theories in psychology and education research domains regarding \textit{mindsets}, e.g. growth versus fixed mindsets~\cite{Bernecker2019} and mindsets related to phases of action~\cite{Gollwitzer2012}. The study of mindsets is very useful in providing appropriate recommendations to users, for example, to points of interest in a city~\cite{Viswanathan2022}.

In software engineering, the term mindset is associated with engineering practices or tasks. For example, the Agile mindset is a common terminology used to describe the principles required to practice agile software development, e.g. flexibly adapting to change instead of rigorous prior planning~\cite{Russo2021}. To \textit{define} what the Agile mindset involved, Mordi and Schoop analysed 23 papers published in research venues or privately by practitioners, and performed interviews with 17 industry practitioners~\cite{Mordi2020}. They derived 27 characteristics to describe the ``Agile Mindset''. For example, they reported that all sources had evidence of \textit{trust} and \textit{responsibility and ownership} as characteristics of the mindset. Motogna et al. conducted a study with 47 student teams in a 14-week long course, in which students were encouraged to follow Agile practices~\cite{Motogna2021}. The authors reported that developing appropriate soft skills associated with such characteristics is important to help adopt the mindset needed.

Software engineering research has also focused on the \textit{privacy mindset}, which describes the thoughts of people as they navigate privacy features and invasions in online websites~\cite{Hadar2018, Iwaya2023}. Similarly, the exploration of the \textit{security mindset} describes how people think about security in source code and software~\cite{Tahaei2021}. To understand security and privacy mindsets, Arizon-Peretz et al. conducted interviews with 27 practitioners and analyzed their responses, in the context of themes from organizational climate theory, i.e. factors of the working environment~\cite{ArizonPeretz2021}. Despite extensive research on mindsets, there is no common definition for the term \textit{mindset}~\cite{Buchanan2024}, as it depends on the perspective and context of the related research. In this paper, we identify documentor \textit{mindsets}, which capture the implicit relations between considerations across the different dimensions of the documentation creation process, based on interviews with 26 documentation contributors.

\section{Study Design}
\label{sec:study_design}

We conducted semi-structured interviews, which we subsequently analysed using card sorting~\cite{Hudson2013} to gain a better understanding of the software documentation contribution process.

\subsection{Informant Recruitment}

We directly invited documentors, instead of having an open call for participation, to ensure that informants had regularly and recently contributed documentation. We focused on people who released blog articles or YouTube videos about some software technology. Although we began our search through documentation for Java and Python, we did not disregard documentors of other technologies. Popular blogging websites such as \href{https://medium.com}{medium.com} and \href{https://hashnode.dev}{hashnode.dev} do not provide a standard method to contact bloggers, and contact information such as email was rarely provided. Instead, we recruited the first informant via personal contacts, and used different techniques to subsequently identify documentors:
\\
\\
\noindent \textbf{Github:} For each of Java and Python, we used the Github API to retrieve repositories that were in English, and contained both the name of the technology and the word `tutorial' in either the name, description, or README of the repository. To recruit more informants, we further expanded our query to C++, Ruby, and SQL.
\\
\\
\textbf{YouTube:} For each of Java and Python, we manually searched for the following queries in the search engine DuckDuckGo, in the video tab, in incognito Chrome browser:
\begin{enumerate}
    \item $<$technology$>$ tutorial
    \item $<$technology$>$ programming tutorial
    \item $<$technology$>$ development tutorial
\end{enumerate}
and retrieved each of the search results from the first three pages of the queries.\footnote{We used the common term `tutorial' to identify instructional resources, as opposed to other forms of documentation such as reference documentation.} In total, we obtained 409 unique video links for Java and 421 unique video links for Python.
\\
\\
\noindent For each of the Github and YouTube search results, one author manually determined if the contributor was an individual, i.e. not a community of creators or a company. The author also confirmed that the contributor had contributed documentation related to the working and usage of a software technology, irrespective of the technology they were documenting, within the past six months, and at least a total of three times. 
\\
\\
\noindent \textbf{WriteTheDocs:} WriteTheDocs is ``a global community of people who care about documentation''~\cite{writethedocs}. The community has a Slack workspace in which technical bloggers can introduce themselves in the Slack channel \texttt{intros}, and share their work in the channel \texttt{community-showcase}. Between January and April 2023, we monitored both of these channels and invited bloggers who had created a post in the past two months about the working and usage of software, and had created at least three blog posts so far.
\\
\\
We obtained the publicly available email IDs of the identified contributors and sent them an email inviting them for an interview. Table~\ref{tab:informants} shows the details of the 26 informants we recruited.\footnote{We interviewed two additional documentation creators, recruited via references of two informants. However, one creator created documentation for their own software in the form of linux man pages, and the other created documentation for online courses that were not publicly available. We do not report on these two interviews, as they are out of scope of our focus of publicly-available online software documentation.} Informants were not monetarily compensated for participating in the study. The study protocol was approved by the Ethics Review Board of McGill University.

\begin{table*}[htbp]
\centering
  \caption{Details of the informants of the interview study, all of whom are documentors.}
  \label{tab:informants}
  \resizebox{0.9\textwidth}{!}{%
    \begin{threeparttable}
  \begin{tabular}{cllrrl}
  \toprule
  & \textbf{Recruited from} & \textbf{Type of content} & \textbf{Programming exp. (yrs)} & \textbf{Documentation exp (yrs)} & \textbf{Familiar technologies} \\
  \midrule
  P1 & Reference & Text & 25 & 13 & Javascript, Python \\
  P2 & Github & Text & 40 & 10 & C++, Python \\
  P3 & Github  & Text \& Video & 25 & 6 & Java, Kubernetes \\
  P4 & Github & Text & 33 & 9 & Python \\
  P5 & WriteTheDocs & Text & 5 & 1 & Python \\
  P6 & Github & Text & 24 & 1 & Python \\
  P7 & Github & Text & 1 & 0$^*$ & NodeJS, Cloud \\
  P8 & Github & Text \& Video & 23 & 3 & Python, Rust \\
  P9 & WriteTheDocs & Text & 4 & 2 & Javascript \\
  P10 & YouTube & Text \& Video & 22 & 17 & Java, Spring \\
  P11 & YouTube & Video & 5 & 3 & Python, Plotly, Dash \\
  P12 & Github & Text & 21 & 13 & Asp.net, C\#, HTML \\
  P13 & WriteTheDocs & Text & 3 & 2 & NodeJS, ReactJS \\
  P14 & WriteTheDocs & Video & 5 & 3 & Git \\
  P15 & Github & Text & 20 & 17 & PHP, Javascript, Python, Go \\
  P16 & Github & Text & 16 & 9 & HTML, CSS, Javascript \\
  P17 & YouTube & Video & 10 & 5 & Python \\
  P18 & YouTube & Video & 8 & 5 & Java \\
  P19 & WriteTheDocs & Text & 9 & 2 & Python, Docker, Git \\
  P20 & Github & Text & 9 & 4 & SQL, C++, Python \\
  P21 & Github & Text & 12 & 4 & Python, GNU/Linux \\
  P22 & Github & Text & 2 & 0$^*$ & Python, C++
  \\
  P23 & Github & Text & 8 & 4 & Javascript, Typescript, NodeJS
  \\
  P24 & Github & Text & 25 & 10 & Javascript, Typescript, web dev.
  \\
  P25 & YouTube & Video & 12 & 3 & C++, Java \\
  P26 & Github & Text & 10 & 7 & C++, Python \\
  \bottomrule
  \end{tabular}
  \begin{tablenotes}[online]\footnotesize
\def\tnote#1{\protect\TPToverlap{\TPTtagStyle{#1}}}%
\item[+] If the informant did not self-report the extent of their experience during the interview, we retrieved this information from their LinkedIn profile or public documentation. The reported experience is as of when we interviewed the informant, and is rounded to the nearest year.
\item[*] 0 indicates that the informant's experience was less than six months at the time of the interview.
\end{tablenotes}

  \end{threeparttable}
  }
\end{table*}

\subsection{Data Collection}

Since our goal was to understand \textit{why} and \textit{how} people contribute software documentation, we performed semi-structured interviews~\cite{Lazar2017}, as opposed to structured interviews based on a pre-defined set of questions. This approach gave us the opportunity to gain context-specific insight based on the informant's experiences. We asked informants about their journey into documentation and their process of creating documentation, including how they selected topics, and what their procedure for creation was. We asked informants to expand upon aspects that they would bring up, allowing them to steer the conversation according to their documentation experience. We incorporated any new aspects, such as whether and how they announced their newly released documentation, in interviews with later informants. 

\subsection{Qualitative Analysis}

We qualitatively analysed the interviews in a manual, iterative manner. After completing the interviews with the first five informants, the first author began open coding the transcripts. This process resulted in identifying three major dimensions of the documentation process. \textit{Motivations} describe the incentives for contributing documentation. \textit{Topic selection techniques} capture how to decide on topics to create documentation about. \textit{Styling objectives} describe the purpose of and rationale behind content and presentation design decisions.

The three dimensions occur differently during the documentation process. Motivations act upon the documentor and \textit{encourage them to contribute documentation}. In contrast, \textit{topic selection techniques} and \textit{styling objectives} are dimensions that the documentor decides upon during the documentation process. Thus, the documentor \textit{designs their documentation through} these two dimensions. We use \mots for motivations, \ts for topic selection techniques and \so for styling objectives, to indicate when and how each dimension manifests during documentation contribution.

For each dimension, one author performed card sorting~\cite{Hudson2013} of the open-coded responses from the informants, to identify documentors' \textit{considerations} during the documentation process. Then, all the authors together discussed the relevance, meaningfulness, and accuracy of the identified \textit{considerations}. We identified five motivations,\footnote{We reported on the motivations in a previous short paper~\cite{Arya2024}.} five topic selection techniques, and six styling objectives. We describe documentors' considerations across the three dimensions, in detail, in Section~\ref{sec:considerations}.

\subsection{Mindset Elicitation}
\label{subsec:mindset_eduction}

\begin{figure*}
    \centering
    \includegraphics[width=.95\linewidth, trim=0 60 0 0,clip]{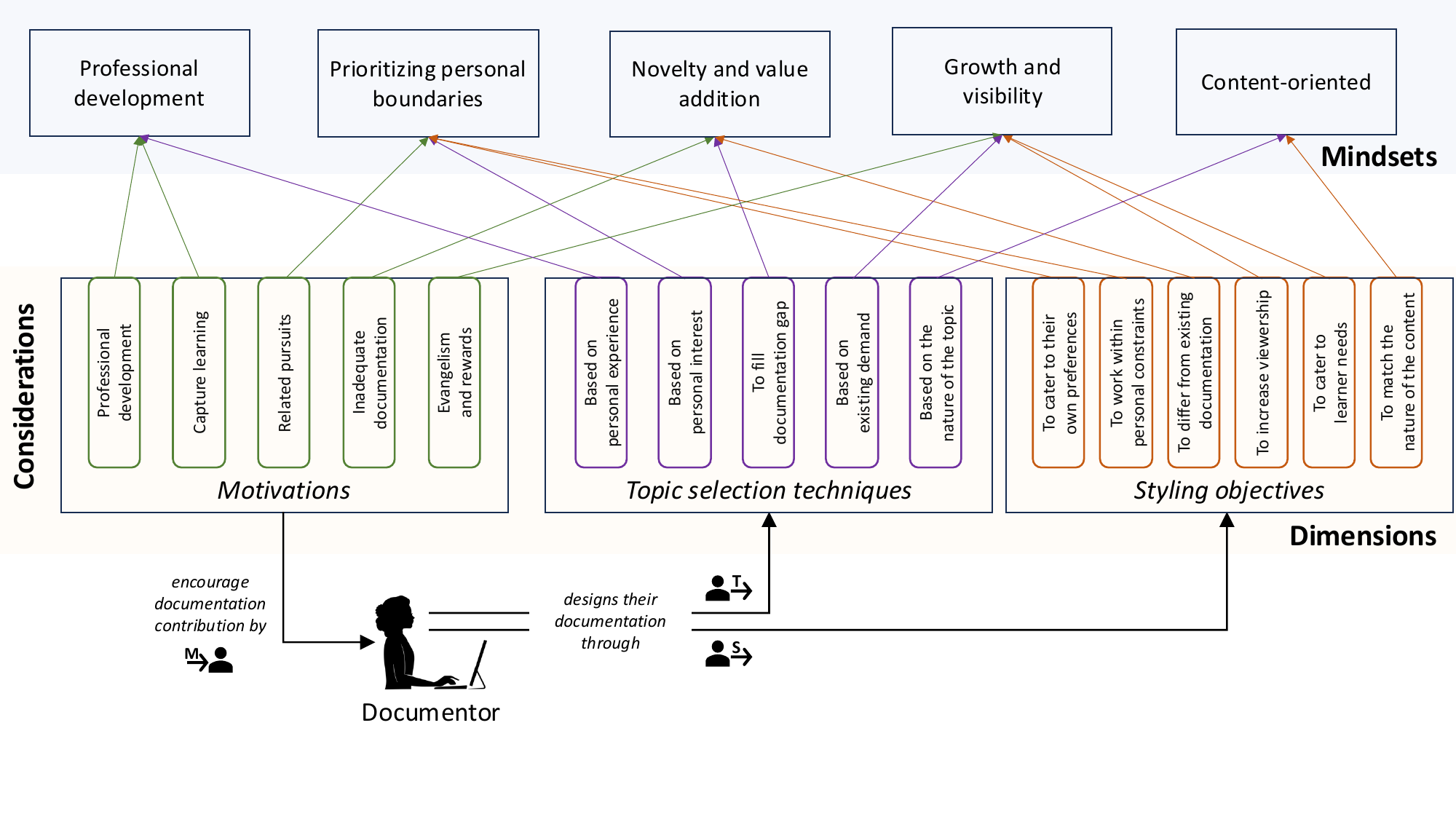}
    \caption{Framework of documentors' mindsets and their associated considerations across the three dimensions of the documentation contribution process, i.e. \textit{motivations}, \textit{topic selection techniques}, and \textit{styling objectives}, based on interviews with 26 documentors.}
    \label{fig:mindsets}
\end{figure*}

We noted that different considerations can be thematically related across dimensions. For example, P3 described their motivation to contribute documentation: \qq{I wanted my examples to be available for everybody to be used and that's why I started putting content on GitHub.}{P3} They also described that they selected topics based on popularity: \qq{a lot of people should be interested in that topic.}{P3} However, we could not reliably infer that P3 selected popular topics \textit{because} they wanted their examples to be used by more people. Still, P3's motivation and topic selection technique are both related to reaching an audience. We grouped such related considerations and thereby elicited five \textit{software documentor mindsets}, that provide an overview of what documentors think about when contributing documentation. We propose the mindsets and their associated considerations across three dimensions as a framework for understanding why and how people contribute documentation, as seen in Figure~\ref{fig:mindsets}. We describe the mindsets, in detail, in Section~\ref{sec:mindsets}.

\subsection{Validation}

We performed member checking to determine whether the mindsets credibly capture informants' thought processes during the documentation process~\cite{Seaman1999, Creswell2014}. We sent informants a questionnaire with a preliminary, one-line, description of the mindsets. We asked respondents to select to what extent they agreed to having experienced that mindset, on a four-point agreement scale with an additional ``Unsure'' option. If they agreed, we asked them to briefly describe how they experienced the mindset. At the end of the questionnaire, we asked respondents if they had experienced any mindsets that were not captured by the five we elicited, and whether they had any additional comments. Whereas one email invitation bounced, we received a total of 17 responses, which we discuss in Section~\ref{sec:discussion}. Further details about the study, including the interview guide, the open-coded data, and the validation questionnaire is available in the accompanying online artifact.\footnote{\url{https://doi.org/10.5281/zenodo.14416777}}

\subsection{Trade-offs}

Our informant pool is a convenience sample of identified documentors whose contact information was publicly available and whose documentation was freely accessible. The limited number of informants is a consequence of conducting lengthy interviews. We chose to accept the trade-off~\cite{Robillard2024} of fewer participants from interviews, as opposed to potentially more participants from a survey study, in favor of obtaining deeper insights on the documentation process. Any novel observations in future work, of the documentation process in alternate contexts, such as that of paid documentation creation, can be integrated into the proposed framework in Figure~\ref{fig:mindsets}.

The interview methodology relies on reflection and self-reporting, wherein informants must recall and describe their own experiences. There is a risk of differences between perceived and actual documentation contribution behaviour. An alternative would have been to conduct an observational study in a natural or lab setting. However, in a natural setting, observing and analysing the documentation process is not feasible, as documentation contribution does not necessarily have a set time frame. Alternatively, developing a simulated setting for a lab study would remove the important context of \textit{volunteered} contributed documentation. Thus, we decided to conduct interviews. We followed a specific-to-general interview technique~\cite{Hove2005} in which we asked informants to explain their thoughts and actions with concrete examples and then followed up with a question on whether this was their general procedure. This technique helped provide context and evidence for the interview responses.
\section{Dimensions of the Software Documentation Contribution Process}
\label{sec:considerations}

Our analysis of the interviews revealed \textit{considerations} informants had while contributing documentation, across three \textit{dimensions} of the documentation process: \textit{motivations}, \textit{topic selection techniques}, and \textit{styling objectives}.

\begin{table*}[tbp]
    \centering
    \caption{Documentors' considerations along the dimension \textit{motivation}.}
    \resizebox{\textwidth}{!}{%
    \begin{tabular}{p{4cm}p{9cm}p{4.5cm}}
         \toprule
         \textbf{Consideration} & \textbf{Description} & \textbf{Example open code} \\
         \midrule
         Professional development & Informants described that they were contributing documentation as a portfolio to demonstrate their knowledge to potential employers and customers. & Good marketing (of self) as a freelance consultant \\
         & \\
         Capture learning & Informants created documentation about what they were learning, to understand the technical aspects of software better. The documentation also acted as a repository of information that the informant could refer to in the future. & To understand the topics, by explaining to "someone else" via the documentation \\
         & \\
         Related pursuits & Informants created documentation because they were curious about what it involved, or because they had related interests such as teaching. & Connected passion for writing with technical knowledge as "own art" \\
         & \\
         Inadequate current documentation & Informants created documentation to overcome the issues they faced with documentation when learning or searching about technical topics. & Existing documentation has too many details [for a beginner] \\
         & \\
         Evangelism and rewards & Informants created documentation to help others (altruism) or to gain benefits (e.g. monetary compensation). & Put up [documentation] content to get minimum level of income \\
         & \\
         Other motivations & Informants described other motivations, such as being inspired by authoritative people to contribute documentation, or wanting to receive feedback. & Tutor at boot camp mentioned technical writing so decided to try it \\
         \bottomrule
    \end{tabular}
}
    \label{tab:motivations}
\end{table*}

\subsection{\mots Motivations}

We elicited five major motivations that capture why our informants contributed software documentation (see Table~\ref{tab:motivations}).

\subsubsection*{\textbf{Professional development}}

Informants expressed the importance of their documentation as an online portfolio of their knowledge for potential recruiters (13 informants). For example, documentation provided an opportunity for informants to keep up with technologies of interest that they were not exposed to in their regular employment (P23). Informants also received external incentives to contribute documentation, such as encouragement from their employers (P1, P10, P15, P20). When the informant represented the company publicly, the created documentation provided evidence of their authority: \qq{We're also encouraged to have our own brand and my boss wants me to be out there promoting my brand so that when we're at conferences, people look to me as an authority, like this person does know what they're talking about.}{P10} 

\subsubsection*{\textbf{Capture learning}}

Informants described that their documentation was like a systematic set of notes (P8, P15, P25), which they could refer to in the future (P16, P21): \qq{If you have to work hard to find something out by pulling lots of things together, then putting it in one place in your blog is cool. Then the next time I can use those instructions. So it's just notes, but public.}{P15} This was prompted by frustrations of forgetting useful information (P16, P24): \qq{I wanted to have an archive for myself that I could consult. I had the experience where I would run into some problem, spend three hours fixing it, and then I would move on to the next thing. A month later, I would have the same problem and I wouldn't remember how I fixed it, so I'd have to spend another three hours rediscovering the solution.}{P16} Creating the documentation was a way to help understand a topic better (eight participants): \qq{When you have to explain something to someone, then whether you can \textit{actually} explain shows that you understand something.}{P5}

\subsubsection*{\textbf{Related pursuits}}

For some informants, creating documentation was the result of combining programming with another personal pursuit, e.g. programming and teaching (P17, P19), video creation (P18, P25), or writing (P2, P4). \qq{I wanted to make a YouTube channel because me and my friends from high school would make YouTube videos for video games. [...] I thought, well I have this programming experience [...] and I ended up deciding to make programming tutorials.}{P18} Documentation also provided a way for informants who had struggled while learning to utilize their knowledge and stay connected to programming (P5, P9). P5 expressed that although they were interested in learning programming, they did not think they would ever become a full-time programmer. Instead, blogging would allow them to still get an idea of how things work. Similarly, P9 explained: \qq{I just felt like: I'm really struggling with coding. I don't feel fulfilled when I write code. What about just trying out technical writing?}{P9}

\subsubsection*{\textbf{Inadequate current documentation}}

When the documentation was lacking (P12, P19), inaccessible to beginners (P4), overwhelming (P5), or scattered across multiple resources (P23), informants felt the need to fill the gap with relevant documentation. P12 explained: \qq{That's really how I ended up writing: I found that an awful lot of the documentation online was either nonexistent or quite obtuse. It was difficult to read from a beginner point of view. So I try things and then write an article about it and put it on my blog.}{P12} Informants described that existing documentation did not cater to their preferences~\cite{Arya2022}, prompting them to create pertinent documentation: \qq{I figured that a lot of people would appreciate recorded video material because that's what I appreciated. And I saw a gap there.}{P11} Informants could then tailor their documentation to their needs: \qq{[I decided] I will start a blog where I will write content the way I wanted people to write them, when I was still learning.}{P23}

\subsubsection*{\textbf{Evangelism and Rewards}}

Informants felt the need to ``give back''~\cite{MacLeod2015}: \qq{I had learned a ton of stuff on YouTube throughout my educational journey [...] So I was like - I'm going to teach people on YouTube.}{P17} They were motivated to help other people gain access to knowledge: \qq{I wanted my [code] examples to be available for everybody and that's why I started putting content on GitHub and wherever possible writing articles, so other people can access the content.}{P3} Although monetary compensation can drive documentation creation (nine informants), informants explained that it did not always reach their expectations (P8). Yet, other benefits such as positive comments from users were motivating (seven informants): \qq{[When] I started the YouTube channel, the videos did receive some good feedback and I just decided to continue with a few more topics that I liked. And then this was all coming together and I started doing more and more videos on the channel.}{P25}

\begin{table*}[tbp]
    \centering
    \caption{Documentors' considerations along the dimension \textit{topic selection technique}.}
    \resizebox{\textwidth}{!}{%
    \begin{tabular}{p{3cm}p{9.5cm}p{5cm}}
         \toprule
         \textbf{Consideration} & \textbf{Description} & \textbf{Example open code} \\
         \midrule
         Based on personal experience & Informants decided to write about a topic that they had experience with, e.g. a topic they worked on at work, or while learning. & Solutions to issues faced at workplace \\
         & \\
         Based on personal interest & Informants selected topics that they were curious about and/or that interest them. & Learning [topics] to support kids' interests \\
         & \\
         To fill documentation gap & Informants selected topics for which they did not find documentation or the existing documentation did not explain the topic well. & If a topic may benefit a lot of people and has not already been covered well on YouTube \\
         & \\
         Based on existing demand & Informants selected topics that many people would want to learn about. & Multiple people asked the same question online, so create blog post to refer them to \\
         & \\
         Based on the nature of the topic & Informants selected topics that have a particular ``documentability'', e.g. they are appropriate for beginners, or are easy to create a tutorial about. & One [topic] that is slightly obscure that can be justified well by an article \\
         & \\
         Undefined & Informants described "naturally occurring" ideas or that they did not know exactly how the ideas for topics came to them. & May come from a variety of ideas - no specific pattern \\
         \bottomrule
    \end{tabular}
  }
    \label{tab:topic_selection}
\end{table*}

\subsubsection*{\textbf{Other motivations}}

Informants described other factors, although less notable than the previous five categories, that motivated them to contribute documentation. For example, someone else advocated for creating documentation (P6, P7, P16, P21): \qq{It was from an article [... the author] mentioned technical writing. And also from my program in a bootcamp, [...] a tutor came and talked about it. [...] That's why I just decided to try and write.}{P7} Informants (P10, P18) wanted \qq{something tangible that I could build that was my own.}{P10} Informants also wanted to build a network (P26) to obtain feedback~\cite{Parnin2013}, and improve other skills like writing and English language skills (P21).

\subsection{\ts Topic selection techniques}

We elicited five techniques informants used to decide what topics to create documentation about (see Table~\ref{tab:topic_selection}).

\subsubsection*{\textbf{Based on personal experience}}
\label{subsubsec:personal_experience}

Informants wrote about topics that they were familiar with, for example, technology that they had recently learned about in their daily work (21 informants): \qq{My day job is in Python, so if I'm doing something useful and it's interesting at work, I may spend an hour in the evening documenting.}{P8} This meant that the informants were already familiar with the topic: \qq{I thought that it could be a nice idea to review all the basics and share the knowledge I already have.}{P22} An advantage of this technique is that the content comes from experience with a real application: \qq{I basically baked down all of the things that I did [at work] and put it into that video. So it was very much real world experience baked into a video.}{P17}

\subsubsection*{\textbf{Based on personal interest}}

The voluntary nature of contributing documentation allowed informants to explore topics that they were curious or passionate about (15 informants): \qq{Most of the time, these were topics that I personally liked and wanted to investigate more. I already worked with micro-controllers, but I had never gone into every single detail of it.}{P25} Contributing documentation provided the opportunity to go beyond their professional experience: \qq{I purposefully picked technologies I would not have any exposure to at my job, but that I'm interested in and I don't want to wait around to have to learn.}{P6}

\subsubsection*{\textbf{To fill documentation gap}}

Informants selected topics for which they found the existing documentation either too complex for beginners, not extensive, not to their preferences, or completely lacking (seven informants): \qq{People were explaining how [something] was working, but they weren't showing how they did it.}{P22} They focused on difficult topics to document, as there would be fewer resources on such topics (P6, P17, P19). Traffic analytics, such as failed searches for what topics people wanted to learn about but did not find (P1), and volume of existing documentation on a particular topic (P2) were helpful to make informed decisions about what topics to cover.

\subsubsection*{\textbf{Based on existing demand}}

Informants selected topics because there was a clear demand from the audience for particular topics. They monitored community channels and question forums (six informants): \qq{I see what people ask because usually I can answer them. Or I can tell them to search my blog or my website and I know there's a solution there. Sometimes the questions come up and I cannot answer them right away, or the answer would be something larger and I have to try it out myself. And that gives me a topic for a blog post.}{P24} Informants also directly asked their audiences what topics they wanted (P9, P10, P20): \qq{In one of the newsletters, I asked them if they want me to cover some specific topic. They can just send me a message about it, and I will do it.}{P23} Additionally, informants also selected topics that, from their experience, they knew people would want (P1, P7, P12), and would get more views (P8, P10, P18, P25).

\subsubsection*{\textbf{Based on the nature of the topic}}

Informants considered how suitable a topic is to documenting, e.g. by catering to particular audiences (P1, P15), such as beginners (P2, P4, P5), or selecting topics that are applied and hands-on (P2, P14, P17, P26). Informants developed topics towards an actionable deliverable (P17, P22): \qq{Selecting the actual topics was like building off of what I had already [documented]. I built my first game tutorial after making five introductory Python lessons that cover the basics to build that game. So, I was building the blocks... Now I can make a game.}{P17} Whether the topic is ``documentable'' is important: \qq{One step should be self-explanatory, so you go step by step. And so that's why I've kind of moved away from those other [topics] that would be complicated.}{P5}

\subsubsection*{\textbf{Undefined}}

For some informants, ideas for topics manifested naturally (P1, P2, P5, P16). \qq{There's not a clear pattern [...] I'll be walking down the street and I'll have an idea for something that I think could make a good blog post. Or even sometimes it is really granular, like an interactive kind of visualization I have in mind for a particular way to describe some particular concept. And so I jot down whatever notes I have into Notion [a note-taking tool]. And after a while, if the same idea keeps coming back to me, I figure it's probably a good idea to actually write about it.}{P16}

\begin{table*}[tbp]
    \centering
    \caption{Documentors' considerations along the dimension \textit{styling objective}.}
    \resizebox{\textwidth}{!}{%
    \begin{tabular}{p{3cm}p{9cm}p{5.5cm}}
         \toprule
         \textbf{Consideration} & \textbf{Description} & \textbf{Example open code} \\
         \midrule
         To cater to their own preferences & Informants incorporated aspects that they would have liked to have as a learner. & Is participant's personal preference to have a structure with situation, problem, and solution \\
         & \\
         To work within personal constraints & Informants styled their content based on how much time they had, or to keep an achievable routine for posting documentation. & Code has to be short enough that it can be written about in a relatively short time \\
         & \\
         To differ from existing documentation & Informants took deliberate measures to ensure their documentation was different from existing documentation about the same topic. & Differentiate from existing static blogs by providing interactive real examples \\
         & \\
         To increase viewership & Informants used techniques to capture a learner's attention and optimize for search engines. & Keep reader engaged and incentivized to read until the end, by adding humor \\
         & \\
         To cater to learner needs & Informants styled content to what they thought a learner would want or need. & Have three examples so at least one may be relevant / valuable to the viewer \\
         & \\
         To match the nature of the content & Informants styled content based on what suited it best, e.g. examples are well suited to explaining hands-on topics. & GUI software lends itself better to screenshots/images \\
         & \\
         Other objectives & Other styling objectives that are less notable. & Follow mentor's documentation as a guide \\
         \bottomrule
    \end{tabular}
  }
\label{tab:styling_rationale}
\end{table*}

\subsection{\so Styling objectives}
\label{subsec:styling_rationale}

We elicited six objectives that informants had for the organization and presentation of content (see Table~\ref{tab:styling_rationale}).

\subsubsection*{\textbf{To cater to their own preferences}}

Informants created documentation the way they like documentation to be (seven informants):
\qq{I actually don't really enjoy reading that much, because there are a lot of words. I like reading simple things, so this is why I try to make [my documentation] simple.}{P7} Informants acknowledged that they focused on what worked best for themselves: \qq{[In my documentation,] I always like to follow [the structure:] the situation, the problem, and the solution. Because for me - and I'm definitely biased - this is the way I learn.}{P23}

\subsubsection*{\textbf{To work within personal constraints}}

Informants described how they styled their content based on what was convenient and feasible to their time and effort (P2, P4, P11, P14, P18, P21). For example, to determine what code examples to add, P2 described: \qq{It has to be code that's simple enough that I can write about it in a relatively short time.}{P2} Informants modified their documentation accordingly: \qq{If it's taking me too long, I just try my best to wrap it up, don't put anything else. Just put a link to the original articles or some Stack Overflow questions and answers and finish it.}{P21}

\subsubsection*{\textbf{To differ from existing documentation}}

Informants strategically styled their documentation to stand out from existing documentation (P8, P16, P17): \qq{There's a thing called [technology name]. It's basically a wrapper where you can paste in your code and then it creates permalink. I put the permalink in my video description. Not everybody does that in their video, so I'm hoping that makes it a nice feature of my channel.}{P8} Similarly, P16 added interactive components in their textual blog, to dynamically see the impact of code changes: \qq{Most blogs have some sort of static format like Markdown. [...] So the goal with my blog was to be able to create these one off components [...] that can be then embedded in the blog post.}{P16}

\subsubsection*{\textbf{To increase viewership}}

When designing their documentation, informants added features that would capture the attention of their audience (nine informants):
\qq{Your [article] title is really really important. Because at the end of the day, you want to have a title that when someone searches it on Google, it might pop up.}{P9} Additionally, they optimized for search engines to index the documentation on earlier pages (P1, P9, P18, P23, P25): \qq{I'm also trying to add the YouTube chapter markers to videos. YouTube is promoting videos better if there are chapter markers in there.}{P25}

\subsubsection*{\textbf{To cater to learner needs}}

Informants were cognizant of the information and styling needs of users when designing their content. For example, documentation needs to be very clear and concise for the audience (nine informants): \qq{In the software world, developers follow some technical jargon. I avoid using them because I need to explain to an audience and it should be really clear.}{P19} Informants deliberately styled their content to help the audience understand better (ten informants): \qq{If you are new to this area and some people are explaining things really, really quickly and you just have a big chunk of a code snippet at the end, [...] you are just like: OK, so what does this do, why this, why that. That's why, [in my documentation,] every single step has a code snippet rather than just one at the end.}{P5} To assist the audience, informants provided useful indicators about the content (P13, P16, P20, P24, P26), such as a note about prerequisite information requirements: \qq{[The readers] need some kind of knowledge before, to better understand the article. So I always make sure I let them know what they need.}{P13}

\subsubsection*{\textbf{To match the nature of the content}}

In some cases, the content is naturally well suited to a particular type of presentation (P1, P3, P4, P15, P17), and the informants leveraged this characteristic when creating their documentation. \qq{The type of software that I work on mostly tends to be data visualization or data analysis software, which lends itself very well to nice canned examples that go from top to bottom.}{P1} Informants styled their content based on whether they wanted to cover a breadth of topics or wanted to go in-depth: \qq{Over an one-hour time period, I can cover a huge range of topics [by providing hands-on examples], which would not be possible if I [document] in a pictorial way.}{P3} Informants covered the most necessary details to make their documentation self-contained (P24), and provided an appendix (P21) or links to additional material (P15, P25), to avoid overwhelming readers.

\subsubsection*{\textbf{Other objectives}}

Informants described other reasons and goals for styling. For example, informants followed best practices: \qq{A basic concept in technical writing is [covering] a single topic. So you should stick to one thing [in a single document].}{P5} They were also inspired by other resources: P16 saw other documentation have a note with the intended audience, and incorporated the idea in their own documentation because they found it valuable.

\section{Software Documentor Mindsets}
\label{sec:mindsets}

We observed that different considerations can be related by common, implicit themes. For example, the \mots motivation \textit{lack of inadequate documentation}, the \ts topic selection technique \textit{to fill documentation gap}, and the \so styling objective \textit{to differ from existing documentation} are all related to contributing new and improved documentation. The first author identified and grouped considerations by the underlying themes. We obtained a total of five groups, which all the authors discussed based on evidence from the interviews. The groups capture what documentors think about during the documentation creation process, across the three dimensions. Thus, we identify them as \textit{software documentor mindsets}.

Figure~\ref{fig:mindsets} shows the framework of the five mindsets observed among our informants, and the corresponding considerations across the three dimensions of the documentation process. These elements of the framework impact documentation creation differently. \mots \textit{Motivations} encourage people to contribute documentation, whereas the documentor designs their documentation through \ts \textit{topic selection techniques} and \so \textit{styling objectives}. We mapped the mindsets for each informant, by identifying whether the informant described the associated considerations. The mapping, a sample of which is shown in Table~\ref{tab:observations_per_informant}, documents how the mindsets manifest for each informant. For example, for P5 we identified multiple considerations associated with the \textit{personal development} mindset. In contrast, P1 had this mindset when they were initially motivated to contribute documentation, but more prominently had the mindset of \textit{growth and visibility} when selecting topics and styling their documentation. We describe the five mindsets with examples from our interviews.

\begin{table*}[tbp!]
    \centering
    \caption{Documentors' mindsets and the corresponding considerations across the three dimensions of the documentation contribution process, for each informant. The complete table for all 26 informants is available in our accompanying appendix.}
    \resizebox{0.9\textwidth}{!}{%
    \begin{tabular}{lll|lllll}
         \toprule
         \textbf{Mindset} & \textbf{Dimension} & \textbf{Consideration} & \textbf{P1} & \textbf{P2} & \textbf{P3} & \textbf{P4} & \textbf{P5 ...} \\
         \midrule
         \textbf{Personal development} & \mots Motivation & Professional development & X & X & & X & X \\
            & \mots Motivation & Capture learning &  &  & &  & X \\
            & \ts Topic selection techniques & Based on personal experience &  &  & X & X & X \\
         & \\
         \textbf{Prioritizing personal boundaries} & \mots Motivation & Related pursuits &  & X & X & & X \\
             & \ts Topic selection techniques & Based on personal interest &  & X & X & X & X \\
             & \so Styling objectives & To cater to their own preferences & X &  & X & X &  \\
             & \so Styling objectives & To work within personal constraints &  & X & & X &  \\
         & \\
         \textbf{Novelty and value addition} & \mots Motivation & Inadequate current documentation &  &  & X & X & \\
             & \ts Topic selection techniques & Filling documentation gap & X & X & & & X \\
             & \ts Styling objectives & To differ from existing documentation &  &  &  &  &  \\
        & \\
        \textbf{Growth and visibility} & \mots Motivation & Evangelism and rewards &  & X & X & X & X \\
             & \ts Topic selection techniques & Based on existing demand & X & X & X & & \\
             & \so Styling objectives & To increase viewership & X &  & X & X & \\
             & \so Styling objectives & To cater to learner needs & X &  &  & X & X \\
        & \\
        \textbf{Content-oriented} & \ts Topic selection techniques & Based on the nature of the topic & X & X &  & X & X \\
             & \so Styling objectives & To match the nature of the content & X &  & X & X & \\
         \bottomrule
    \end{tabular}
  }
\label{tab:observations_per_informant}
\end{table*}

\subsection*{Personal development}

A documentor having the \textit{personal development} mindset focuses on how their contributed documentation can be used to improve their own knowledge and opportunities. Documentors contribute documentation for their \mots \textit{professional development} or to \mots \textit{capture their learning}, and select topics to cover \ts \textit{based on their personal experiences}.

Learning through teaching~\cite{Sherin2002, Cortese2005} and public note-taking~\cite{Simon2008, Huang2021} are beneficial for both the creators and the audience. Kim et al.'s theory of learning describes that information gain follows three learning stages: the first has mainly declarative knowledge, the second has a mix of declarative and procedural knowledge, and the third has mainly procedural knowledge~\cite{Kim2011}. Documentation contribution follows a similar set of stages in which documentors first gather information, try and apply it for themselves, and document that applied knowledge for others to use. In doing so, documentors are able to learn and retain technical knowledge and accompanying soft skills such as communication.

Thus, although the effort required to contribute documentation is high~\cite{Shmerlin2015}, documentors value the experience they gain: \qq{Honestly, I hate writing [... but] I'm writing [the documentation] for myself, because I forget things and if I cannot clearly express myself, it's going to be very hard for me to understand it later.}{P24} This experience is key for improving the documentor's own professional qualities: \qq{That ability to summarise, communicate, write it down [...] You have to communicate clearly and well in GitHub issues, comments and reviews. So the fact that I write constantly makes me a better engineer.}{P15} With this mindset, documentors primarily document for themselves, and do so by imagining themselves as the information consumer: \qq{I picture my own days as someone who has limited knowledge about what I'm writing about.}{P13} This system of conversing with the ``other self'' who consumes and responds to information that the ``self'' creates~\cite{Murray1982} can help provide clear information: \qq{I'm trying to be explicit so that people understand it without getting confused, because I know that I'll be confused. If I go back to the blog post that I've written in the beginning, I hope I can understand what I've done there.}{P24}

\subsection*{Prioritizing personal boundaries}

As volunteer contributors, documentors have the liberty to mould their documentation to their own constraints and interests, and thus may have the \textit{prioritizing personal boundaries} mindset. This mindset involves creating documentation because of \mots \textit{related pursuits}, selecting topics \ts \textit{based on personal interest}, and styling content \so \textit{to work within personal constraints} and \so \textit{to cater to their own preferences}.

Sansone and Smith defined intrinsic motivation as occurring for individuals when ``their behaviour is motivated by the actually, anticipated, or sought experience of interest''~\cite{Sansone2000}. They further described that to pursue long-term goals, individuals must self-regulate their behaviour to continue to be motivated to experience interest, so that they can reach their goals. With the \textit{prioritizing personal boundaries} mindset, documentors pursue topics of their interest, and design their documentation with styles associated with their preferences. As a result, an implicit self-regulation to retain the motivation to contribute documentation exists. Thus, the documentor's interest is important for documenting. For example, despite the demand for particular topics, documentors can make decisions about what they would like to cover in their documentation: \qq{I just don't find testing very interesting. So none of my content is about it and I don't feel like I'm the best person to teach it, as a result.}{P16} This interest can also act as a proxy for what would engage their audience: \qq{If something is really, really interesting for me, that's a good candidate for what would probably interest all my audience.}{P3}

As voluntary contributors with other commitments, documentors are also bound by the amount of time they have. Documenting familiar content allows documentors to save time because they already have the material they need ready, either in the form of notes (P15, P17, P21) or code: \qq{[The documentation] doesn't exist until the demo is built. If I do a cool thing and think oh, I should write about that, then [the demo] is already done. Whereas if I wanted to write a blog post about something [I have not done], I would need to figure it out and then write the words. But typically the blogging process starts when I've got a demo and all I need are the words.}{P15} To reduce the time spent in actual documentation creation, documentors can allow ideas to first develop in their mind: \qq{As I'm coding, I'm thinking about how my writing should be, I'm thinking about the structure. So when I'm actually writing it doesn't take much time.}{P7} Other forms of documentation, such as official reference documentation, are continuously revised to be consistent with the corresponding technology~\cite{Shi2011} to communicate \textit{user-accessible features}~\cite{Dagenais2010}. In contrast, documentors do not have the pressure to create documentation: \qq{I acknowledge that [creating documentation] is really time consuming for me. Sometimes I don't want to do it and then I don't. So I'm not forcing myself to be active and reach an audience because I don't want to make this blow up, I don't want to have a big audience. I just want to teach people that want to learn and I also want to learn.}{P22}

\subsection*{Novelty and value addition}

Documentors having the \textit{novelty and value addition} mindset are conscious of the need to provide documentation with a clear value proposition over existing resources. This mindset stems from experience with \mots \textit{inadequate current documentation}, and involves selecting topics that assist in \ts \textit{filling the documentation gap}, and styling content \so \textit{to differ from existing documentation}.
 
Existing documentation often has many issues, including lack of readability~\cite{Aghajani2019} and irrelevance to real world examples~\cite{Guo2017}, that can hamper the learning of technical concepts~\cite{Robillard2011}. Some of these issues may arise because of the \textit{expert blind spot}~\cite{Nathan2001}, wherein experts, who create the technology and the corresponding documentation, make incorrect assumptions about novices' knowledge and understanding when communicating technical details. Instead, users and other participants not at the core of a community can provide novel perspectives and ideas, as opposed to core contributors~\cite{Safadi2020}. The \textit{novelty and value addition} mindset exemplifies the emphasis that documentors, who are not necessarily directly associated with a software technology, put on how documentation can be valuable. Documentors think about how to contribute unique content: \qq{I was looking for ideas and .Net7 has come out, C\#11 came out. [...] All the other bloggers are going to pick: \textit{here are the new features in C\#11} or they're going to pick topics like \textit{how to write good C}. So I thought, all right, I'm going to choose something totally different.}{P6} Documentors identify how to add value to existing documentation: \qq{Sometimes, I will copy a specific example from W3Tutorials or TutorialsPoint because [...] it’s a simple example. And I try to explain how I would want it to be explained to me, because that’s what they’re missing - the explanation.}{P18}

As users of documentation, documentors recognize the limitation of existing documentation and have ideas to address the struggles of learning from inadequate documentation. For example, a common documentation issue is the lack of real-world examples in documentation~\cite{Guo2017, Aghajani2019}. Documentors who learned software development technology for a specific purpose recognize this issue from their own experiences: \qq{When I was freelancing, I felt that a lot of stuff that I had done earlier on my YouTube channel was theory, but it wasn't real world. And I thought, actually some of this freelance stuff is really real world. And I should make videos on this because these are actual problems and it's not straightforward.}{P8} With the important technical knowledge that links the technology with realistic applications, documentors design their documentation to provide novel, practical examples: \qq{I create a small use case that will help put someone in this problem situation, and then I explain that we will build this [solution] [...] At the end, [the learner] will have a project that they can use in the real world.}{P23} Ultimately, documentation is a communication~\cite{Raglianti2023}, and since different people communicate differently, there is an inherent originality to human-created documentation: \qq{There's this whole thing of: the topic has already been written about, so there's no need to write about it. But, individually, we all have various ways we could add value to people that relate to us, by how we frame our writing.}{P9} As a result, the value and perspective that documentors contribute can not be overshadowed by emerging generative artifical intelligence (AI) technologies: \qq{When you write \textit{create a list for me}, you get like ten points or twenty points [from ChatGPT]. But when you're putting your own words [...] it's like your own art.}{P19} Instead, these systems can be used as tools to support documentors in their creation process~\cite{Bhat2024}: \qq{I think as long as humans are always thinking of ways around problems, we might still just have an advantage over computers and we could just make them do the boring stuff.}{P8}

\subsection*{Growth and visibility}

The \textit{growth and visibility} mindset refers to the pursuit of viewership of the documentation, either to help more people learn, or to reap other rewards such as fame or monetary compensation. Documentors with this mindset are motivated by \mots \textit{evangelism and rewards}, select topics \ts \textit{based on existing demand}, and style their documentation \so \textit{to increase viewership} and \so \textit{to cater to learner needs}.

Kroll introduced three perspectives to thinking about the audience when tailoring written content: \textit{rhetorical}, i.e. the characteristics of an audience, \textit{informational}, i.e. the content they would need, and the \textit{social}, i.e. their attention and preferences~\cite{Kroll1984}. With the \textit{growth and visibility} mindset, analysing what will help engage an audience is part of the documentor's workflow: \qq{you do research on competition versus volume [for a topic].}{P2} Based on their research, documentors take informed decisions: \qq{Ultimately what I'm trying to do is help people. So if 20,000 people are viewing a video and 100 people are viewing a blog post, then it makes sense that Youtube is where this content goes.}{P10} Similarly, styling content to capture a potential audience's attention is critical~\cite{McRoberts2016}: \qq{Maybe they [readers] hit \texttt{ctrl-F} and they search for \textit{cluster}. They don't find it. Boom - we've just lost their attention. So catering to how people will search for things, be it text, be it visual, or the specific words they are using [is important].}{P1} Documentors even prioritize such features during documentation creation: \qq{Titles and thumbnails are very important when it comes to YouTube and click through rate. You could make a great video, but if you are not grabbing someone's attention with both the title and a thumbnail, it's kind of pointless. So I always start there.}{P10} Search engine algorithms retrieve development related information from a variety of platforms, including \url{medium.com} and \url{youtube.com}~\cite{Hora2021}. As a result, documentors take deliberate measures to ensure their documentation is findable~\cite{Bhatti2021-2} and discoverable~\cite{Robillard2011}: \qq{As much as I have the philosophy of writing for humans first before search engines, you also have to consider how your articles will rank on Google, how your article will be searchable, how your article is easily accessible when people try to query that topic.}{P9}

Documentors leverage multiple platforms for their content, as cross-posting can also increase visibility~\cite{Marks2017}. For example, P8 developed three formats simultaneously: \qq{The website was something to show people or redirect people to, when I was applying for jobs. And then it was also a way to accompany the videos, so I was hoping that the website would lead people to the YouTube channel and the YouTube channel would lead people to the website. And all the while I was building up my GitHub repositories for my learning and skills.}{P8} Different formats from an individual documentor complement one another: \qq{I just want a blog post [that points to the video], in case someone is searching, to find the video, because that's where the bulk of the work went into.}{P10} This is especially useful when documentors have paid documentation: \qq{I create the course and then kind of see how I can reuse the content from the course in different ways. So that's when I actually write a lot of articles, create GitHub repositories and also use the course videos to create a lot of [free] YouTube videos.}{P3} Then, the reuse of information has the ulterior motive of guiding audiences towards the paid material: \qq{I wanted to give people a little taste and be like: hey, if my teaching style suits you, then hopefully you buy the course or follow my stuff.}{P14}

\subsection*{Content-oriented}

Documentors having the \textit{content-oriented} mindset allow the environment and circumstances, as well as the content itself, to guide the contribution process. With this mindset, documentors select topics \ts \textit{based on the nature of the topic} and style their documentation \so \textit{to match the nature of the content}.

Prior research has investigated how to present information based on what is to be conveyed to audiences~\cite{Ummelen1996} and general best practices to follow~\cite{Procida_Diataxis}, without discussing the \textit{applicability} of presentation styles to topics and their information. Wright emphasized how the medium, e.g. printed pages versus dynamic web pages, is dependent on the topic at hand, and the medium should be selected based on its suitability to the topic~\cite{Wright1988}. Documentors with the \textit{content-oriented} mindset pay attention to what the content is best suited to, especially because there can be different types of knowledge~\cite{Maalej2013} that can be shared. Characteristics of a topic help make documentors' creation decisions: \qq{I actually started with Seaborn for a few couple of months. And then I discovered Plotly which seemed even more intuitive to me, more user friendly and had more options graphing wise than Seaborn and I liked the Plotly Express documentation more, so that's why I moved to Plotly.}{P11} Similarly, how to style the documentation also depends on the topics and the content that will go into the documentation: \qq{For instance, I have a tutorial [... like] a set of examples for Numpy whose main theme is drawing stuff in Python. And there is not really an order, especially after you are past the first few chapters. You figure out how to make a picture in Python, and then you can grab whether you need circles first or triangles, or some image filters.}{P4} In this case, while the first few sections of the tutorial are sequential to introduce the foundations of drawing in Python, the remaining are modular to allow readers to browse as needed.

It is important for documentors to have thought about the content deeply, even taking a break during the creation process: \qq{It is not okay to come in one day and write the article: draft, edit, and publish. It's really not effective because you have to write and go away from that article, do other things, refresh your mind, and then come back to update the article.}{P9} Documentors with the \textit{content-oriented} mindset think about the long-term relevance of particular information, and thus decide whether it is worth documenting: \qq{Once you put the content out there, things change, but they don't change so quickly that it is irrelevant. So it is just really thinking of things as what will be useful for the next year or two to come.}{P17}
\section{Validation and Discussion}
\label{sec:discussion}

\begin{figure}
    \centering
    \includegraphics[width=\linewidth]{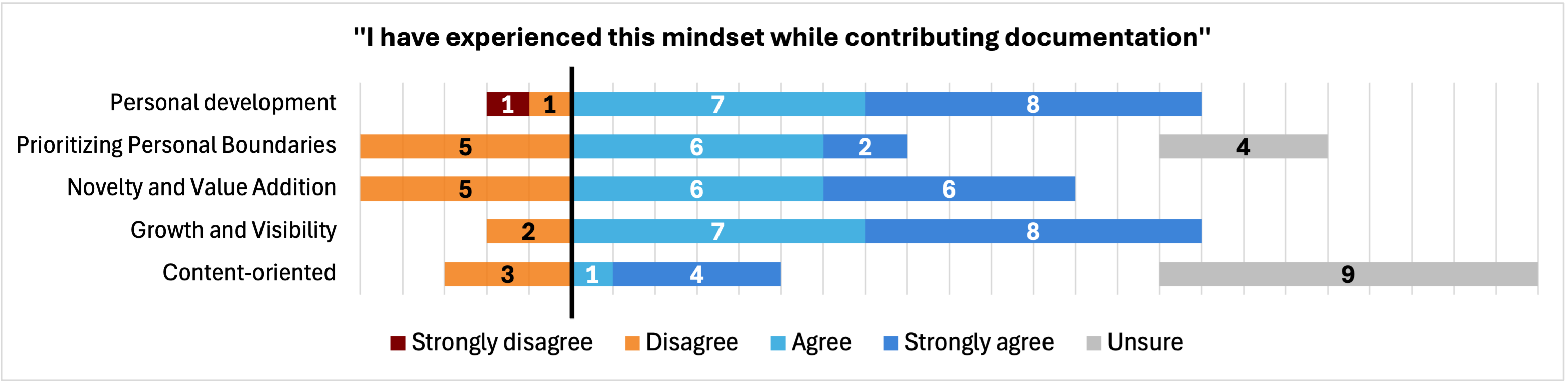}
    \caption{Agreement responses of the 17 respondents to the validation questionnaire.}
    \label{fig:member-checking}
\end{figure}

To determine whether the mindsets we elicited resonate with documentors' experiences, we sent a validation questionnaire to our informants. We received 17 responses, and refer to the respondents as R1-R17, in the remainder of the text.\footnote{The numbers in these pseudonyms do not correspond to the pseudonyms of the informants P1-P17. We do not make the association between respondents and informants to further protect their anonymity.} Figure~\ref{fig:member-checking} shows the responses to the questions with the Likert response format. For each mindset, at least five respondents agreed to having experienced the mindset while contributing documentation. Thus, we can confirm that the mindsets credibly capture the different thought processes during documentation contribution.

We compared the mindsets reported from the questionnaire (perceived) to the mindsets we derived for the same informant (reported) from their interview (i.e. Table~\ref{tab:observations_per_informant}). Although we did not expect to find a perfect correspondence between the two, we noted that there was consistency between the perceived and reported mindsets. For example, R3 ``strongly agreed'' to experiencing the \textit{growth and visibility} and \textit{content-oriented} mindsets. We note that this was also the case in their interviews: they shared anecdotes for many of the considerations associated with these two mindsets. When a respondent \textit{disagreed} with experiencing a mindset in the validation questionnaire, our mappings showed that the informant did not display that mindset dominantly, i.e. they may have discussed a few of the considerations of the mindset during the interview, but not all the associated considerations. Furthermore, the mindsets capture the \textit{latent} attitude and subsequent thought process. One respondent who selected ``strongly agree'' to experiencing the \textit{content-oriented} mindset shared: \qq{This is very insightful and I would not have thought of this myself, but I think you are spot on in identifying this as a mindset.}{R1}

In total, we received thirteen ``Unsure'' responses for the mindsets, with respondents explaining that they did not understand the mindset for four of these cases (one for \textit{prioritizing personal boundaries} and three for the \textit{content-oriented} mindset). We also noted two instances where the mindsets were interpreted differently from our elicitation. For example, R12 ``strongly agreed'' to the \textit{personal development} mindset and explained: \qq{I focus on the reader}{R12}. However, we associate the focus on the external audience with the \textit{growth and visibility} mindset. These confusions may be because we provided a single-line description of each mindset, which can not sufficiently capture the depth of the mindset. We chose to give only a brief description in the validation questionnaire for two reasons. First, we did not want to overwhelm respondents with too much text that could discourage them from completing the questionnaire. Second, we wanted to understand whether respondents' experiences resonated with what we had elicited as a mindset. Had we provided an explanation of the considerations associated with a particular mindset, there was a risk that respondents would consider it a hard constraint on what constitutes the mindset. Responses to the open-ended questions validated that respondents understood the general attitude behind the mindset, despite the one-line description. For example, for the \textit{personal development} mindset, R5 explained: \qq{Resonates with Feynman technique. It helps fill the knowledge-gaps while putting the documentation together.}{R5} This is in-line with our elicitation of the mindset.

The responses to the open-ended questions provided further insight about the respondents' selections. We reflect upon how the mindsets are at play during documentation contribution, and use quotes from the interviews and validation questionnaire as evidence. Specifically, we note how documentors must maintain a balance between multiple mindsets and the challenges they face in pursuing particular considerations. We also discuss mindsets that respondents suggested, many of which provide ideas for future investigation.

\subsection{Balancing multiple mindsets}

The five mindsets we elicited are neither exhaustive nor mutually exclusive, and documentors may display more than one mindset in the creation process: \qq{As a solo content creator, I had to experience most of the mindsets at once, since they all played a factor for me.}{R7} In fact, having a single mindset can even be harmful to the quality of the documentation produced. For example, while being \textit{content-oriented} is beneficial to ensure that the style, topics, and information content go hand-in-hand, it is easy to miss what users may prefer.

Documentors must also think about the visibility of their documentation, to ensure that users are able to find and leverage the information: \qq{We dedicate a bunch of pages to [a topic] and [the documentation pages] don't get used. It's essentially really thinking about our users in terms of what they might want to do and just having examples for that.}{P1} However, focusing only features that promote visibility via search engine optimization (SEO) can interfere with the design of documentation: \qq{Sometimes putting an image [in the documentation] was not really necessary, but I have to, in order to optimize for SEO.}{P23} Such overhead can also disrupt the learning and creation workflow: \qq{[...] sometimes it bores me to create a new post [...] doing the alt tags, putting an image, cropping it, making sure that the key phrase is in there, the SEO stuff. It might take an hour. You know, in that hour I could have learned that [topic] in five minutes and then in the other 55 minutes I could have gone off and learned something way more advanced.}{P8}

Thus, maintaining a fine balance between the mindsets is an integral part of the documentation contribution process. Depending on the goals of the documentor, each mindset may not be weighted equally. For example, R4 and R11 agreed to experiencing the \textit{growth and visibility} mindset, specifying the caveat: \qq{but it is not my top priority}{R4}. The mindsets may also be prioritized differently, based on the documentor's own environment and experiences. One respondent disagreed to having experienced the \textit{novelty and value addition} mindset: \qq{Being too creative with the documentation design could take extra effort, and there is usually no value in doing so.}{R5} Another respondent who disagreed to experiencing the mindset explained that originality occurs naturally: \qq{There might be overlap, but every time someone else is writing about the same thing, it will come out differently.}{R12} In contrast, respondents who agreed to having experienced the mindset emphasized how important it was to intentionally be different: \qq{So much programming content is boring and hard to digest (for me and many of my previous classmates). I knew there was a gap on YouTube for it, and I could fill it.}{R7} We asked respondents to answer the questions based on their own experiences, however, documentors may also recognize mindsets in other documentors. For example, one respondent selected ``Unsure'' for the mindset \textit{prioritizing personal boundaries}, and explained that \qq{I have not experienced this mindset, but I can see how others might adopt it in their work.}{R1}.

\subsection{Challenges with pursuing considerations}

Despite the freedom to prioritize considerations as they like, documentors can become overwhelmed because they must also handle the many tasks related to curating the documentation content, such as editing and audience engagement.

For documentors of video tutorials, creating the video takes additional effort: \qq{I also do all of my own editing and planning. So you can get really lonely. That's what people don't really talk about is that, on YouTube, you're all of these jobs in one. Which is, you know, it's a lot of work.}{P18} Documentors must then \textit{prioritize their personal boundaries} and use multiple strategies to optimize documentation creation. For example, when documenting topics that they already have experience with, the material is ready in advance: \qq{[...] it's something that I've tried and practiced myself and went back and forth, tried something, it failed, tried something else; I keep notes of all that process.}{P21} Similarly, planning by thinking over the content, developing a structure, and writing the code first can help speed up the creation: \qq{Before I write anything, I think: this is how I'll do it. [...] it's kind of stuck in my head for a while and actually writing is the fastest thing.}{P24}

Identifying and connecting with the audience is an integral part of communicating documentation~\cite{Welbourne2016, Bhatti2021}, as it can help information seekers find pertinent resources. Despite being related to the \textit{growth and visibility}, documentors do not have a direct way to communicate with the target audience and understand their actual needs: \qq{I wish I understood better how people are discovering my blog and what they are hoping to get from it.}{P16} They can interact with their audience via webpage comments, or through social media, but this becomes impractical with greater visibility: \qq{there's just so many comments that I'm not going to respond to everyone.}{P17} When audience's needs are difficult to gather, documentors must rely on the \textit{imagined audience}~\cite{Ede1984, Litt2012} and online metrics such as popularity trends of technical topics. However, \qq{You can't really control who is going to read your docs and what they're going to need. You have to look at all of the things that people might need to know and make some arbitrary cutoff decisions.}{P1} Such arbitrary decisions may impact \textit{completeness}~\cite{Aghajani2019}, i.e. how well the documentation informs about a software~\cite{Tang2023}, as not all areas of a technology may be covered~\cite{Parnin2012}. However, documentors, as volunteer documentation contributors, may have mindsets that do not focus on contributing \textit{complete} documentation.

Ultimately, documentation is \qq{[...] a long term project. You really have to be blogging for years before anyone starts paying attention.}{P16} With all the time, effort, and thought spent in contributing documentation, documentors must be cautious of their expectations of return: \qq{you have to try to enjoy the process, enjoy learning new things and trust that it's a long game. It's not - get rich or get famous overnight - type of thing.}{P10}

\subsection{Other mindsets}
\label{subsec:other-mindsets}

Eight respondents answered the question ``Were there any other mindsets that you experienced that we did not ask about?''. Five of these responses are subsumed by our elicited mindsets. For example, R1 and R4 referred to the open source software (OSS) community, where the idea is to ``give back'' and ``empower a greater community'': \qq{Personally, I feel my career and education have benefited a lot from freely accessible content online. Part of the reason I create content (and similar things like contributing to OSS) is to, more or less, return the favor.}{R1} We capture this under \mots \textit{evangelism and rewards}, which is a consideration associated with the \textit{growth and visibility} mindset.

R5 described the role of source code in informing the documentation: \qq{Writing documentation usually makes the author revisit the source code. This might give them ideas about refactoring, or getting a better view of things that should be done, leading to writing code that is more clear, removing bad logic, handling edge-cases, etc.}{R5} R16 also described that the feedback loop \qq{involves regularly revisiting and updating the content based on user feedback, new insights, and changes in the underlying technology or process.}{R16} We suggest further investigation to understand this cycle of code and documentation. Other project stakeholders also influence documentation creation. For example, documenting for colleagues (R13) rather than end users, is helpful to orient teams as well as provide clear insight to businesses about the value of the software (R5). These responses suggest that alternate contexts of software documentation, such as that of collaborative software creation groups and businesses, may introduce new considerations. Our framework of mindsets and their considerations can be expanded to other such contexts of documentation contribution.

R7 suggested the ``empathy mindset'': \qq{Being a college student struggling with Computer Science, feeling that frustration, pain and confusion first-hand was something I knew many others were facing [...] I teach them how I would have wanted programming to be taught to me.}{R7} This respondent captures an integral aspect of the mindsets and the considerations: the \textit{emotions} of the documentor. \mots \textit{Inadequate current documentation} captures the experience of finding a gap in available information that motivates documentation contribution. However, it does not describe the frustration and struggle of experiencing this motivation, that can impact documentation creation. While our focus is on why and how documentors think, future work can explore the emotions of software documentors that contextualize the thought process, and how they impact documentation contribution.
\section{Conclusion}
\label{sec:conclusion}

Despite the availability of existing documentation that accompanies released software, and the effort-intensive task of creating documentation, technology users also curate and contribute documentation online. We interviewed 26 \textit{documentors}, i.e. documentation contributors, who voluntarily created text or video tutorials about software technologies, about \textit{why} and \textit{how} they contribute documentation. We elicited documentors' \textit{considerations} during the documentation contribution process along three \textit{dimensions}: five \textit{motivations}, five \textit{topic selection techniques}, and six \textit{styling objectives}. We grouped related considerations based on their common implicit themes to elicit five \textit{software documentor mindsets}. Our findings surface important insights that impact software documentation research, information seeking and documentation tool support.

\textit{Documentation contribution is a documentor-centric process.} Whereas developers consider documentation as a product that is a supplementary part of a software package~\cite{Rettig1991}, documentors contribute documentation for different reasons, including for themselves. Since documentors have the liberty to make decisions about the documentation they contribute, they can prioritize their own needs, interests, and personal development. Whereas existing software to support the creation of software focuses on the content of documentation~\cite{Azad2017,Head2020}, as well as the needs of the intended audience~\cite{Robillard2017}, documentation creation tools can better support the creation and contribution process by considering the needs and preferences of the documentor.

\textit{As consumers of other documentation, documentors are well-informed to cater to the needs of similarly positioned information seekers.} Prior work has discussed documentation creators as evangelists~\cite{Maher2011}. We introduce a new perspective: volunteer documentation contributors as documentation consumers. Documentors leverage their own learning and programming experiences to inform decisions during documentation creation. As a result, they are informed of the needs of users, an important feature of software documentation production~\cite{Lieberman1991}. Additionally, as the preferred style of learning influences the documentation a documentor creates~\cite{Delanghe2000}, the documentation created can serve audiences who have similar preferences~\cite{EarlePreferences, Arya2022}. This indirect interaction is a partial view of Mehlenbacher's predicted input-output model of documentation creation that suggests that technology creators, documentation writers, and end users are a ``triangle of interrelated technology users''~\cite{Mehlenbacher2002}.

\textit{There is a need to support the management of information consistency across multiple documentation types.} Manually managing information across formats, even to promote visibility of the documentation, can lead to inconsistency~\cite{Arya2020}, due to the duplication of technical information across different documentation types~\cite{Koznov2017}. Prior work has investigated the ability to automatically generate documentation from code comments~\cite{Kramer1999, Head2020}, present information in an alternate manner for code snippets~\cite{Nassif2023}, and support the decision-making judgement of reusing knowledge~\cite{Liu2021}. Our observations suggest the need for a framework to repurpose existing content to maintain traceable links between the different software documentation artifacts. Such a framework can mitigate the risks of information misalignment or inconsistency~\cite{Rahmaoui2019}.

\textit{Contributed documentation introduces the need to revisit documentation quality metrics.} An important aspect of documentation contribution is the documentor's constraints of time and interest. As a result, common standards of documentation design~\cite{Procida_Diataxis, Inzunza2018} and metrics of quality, such as completeness of coverage of information about a technology~\cite{Aghajani2019, Tang2023}, may not apply. Additionally, previously undiscussed challenges arise with contributed documentation. The pursuit of visibility introduces the risk of prioritizing the design of documentation for search engines over human audiences. Furthermore, with increased visibility, amidst an already-involved creation process, engaging with information seekers can become a practical scalability issue~\cite{Zhang2020}.

Our findings suggest the need to support documentors in balancing multiple mindsets: incorporating their own needs, preferences, and learning, contributing original content, reaching the audience, while considering non-human logistics such as suitability of the format to a topic. Consequently, documentors will be able to focus on curating the information that corresponds to their mindsets, and information seekers can systematically identify documentation from documentors who cater to their specific resource needs.

\section*{Acknowledgements}

We thank the informants for their valuable insights about their documentation contribution process. We are especially grateful to the seventeen informants who also responded to our validation questionnaire. This work is funded by the Natural Sciences and Engineering Research Council of Canada (NSERC).

\bibliographystyle{IEEEtran}
\bibliography{9Z_References}

\end{document}